\documentclass[final,3p,times,twocolumn]{elsarticle}
\usepackage{lineno,hyperref}
\usepackage{graphicx}
\usepackage{bm}
\usepackage{subfigure}
\usepackage{amsfonts}
\usepackage[normalem]{ulem}
\usepackage{color}
\modulolinenumbers[5]

\journal{Physics Letters B}

\begin{document}
\begin{frontmatter}
\biboptions{sort&compress}

\title{Evidence that centre vortices underpin dynamical chiral symmetry breaking in $\mathrm{SU}(3)$ gauge theory}

\author{Daniel Trewartha\corref{mycorrespondingauthor}}
\cortext[mycorrespondingauthor]{Corresponding author}
\ead{daniel.trewartha@adelaide.edu.au}
\author{Waseem Kamleh\corref{}}
\author{Derek Leinweber\corref{}}
\address{Centre for the Subatomic Structure of Matter (CSSM), \\
	Department of Physics, University of Adelaide, SA, 5005, Australia}

\begin{abstract}
The link between dynamical chiral symmetry breaking and centre
vortices in the gauge fields of pure $\mathrm{SU}(3)$ gauge theory is studied
using the overlap-fermion quark propagator in Lattice QCD.  Overlap
fermions provide a lattice realisation of chiral symmetry and consequently offer a
unique opportunity to explore the interplay of centre vortices,
instantons and dynamical mass generation.  Simulations are performed
on gauge fields featuring the removal of centre vortices, identified
through gauge transformations maximising the center of the gauge
group.  In contrast to previous results using the staggered-fermion action, the overlap-fermion results illustrate a loss of dynamical chiral symmetry breaking coincident with vortex removal. This
result is linked to the overlap-fermion's sensitivity to the subtle
manner in which instanton degrees of freedom are compromised through
the process of centre vortex removal.
Backgrounds consisting solely of the identified centre vortices are
also investigated.  After smoothing the vortex-only gauge fields, we
observe dynamical mass generation on the vortex-only backgrounds consistent within errors with the original gauge-field ensemble following the same
smoothing.  Through visualizations of the instanton-like degrees of
freedom in the various gauge-field ensembles, we find evidence of a
link between the centre vortex and instanton structure of the vacuum.
While vortex removal destabilizes instanton-like objects under ${\cal
  O}(a^4)$-improved cooling, vortex-only backgrounds provide
gauge-field degrees of freedom sufficient to create instantons upon
cooling.
\end{abstract}


\end{frontmatter}

\section{Introduction}

At the energy scale relevant to everyday matter, Quantum
Chromodynamics (QCD) manifests two key features; the confinement of
quarks inside hadrons, and dynamical chiral symmetry breaking,
associated with the dynamical generation of mass.
Although these phenomena can
be easily shown to exist, the nature of the underlying mechanisms
responsible for them, and whether they share a common origin, have
remained open questions.  

It is generally accepted that these features are both caused by some kind of
topological object in the QCD vacuum which dominates at large distance
scales. Candidates have included objects such as Abelian monopoles \cite{Belavin:1975fg,Polyakov:1976fu,Mandelstam:1974pi}, instantons \cite{Belavin:1975fg,Jackiw:1976pf,Callan:1976je,Polyakov:1976fu,Schafer:1996wv,Trewartha:2013qga} and centre vortices 
\cite{'tHooft:1977hy,'tHooft:1979uj,Cornwall:1979hz,Nielsen:1979xu,Ambjorn:1980ms,Greensite:2014gra,Greensite:2007zz,Bowman:2008qd,Hoellwieser:2014isa,Hollwieser:2013xja,deForcrand:1999ms,Engelhardt:2002qs,Bornyakov:2007fz,Alexandrou:1999vx,Kovalenko:2005rz,Langfeld:2003ev,Ilgenfritz:2007ua,Bowman:2010zr,OMalley:2011aa}.

As the only known first principles technique for studying non-perturbative QCD, the lattice
formulation plays a unique role in studies of these topological
objects. 
In this letter we provide novel lattice QCD results
which reveal a link between centre vortices and dynamical chiral symmetry breaking
in $\mathrm{SU}(3)$ gauge theory.

Centre vortices are topological defects associated with the elementary
centre degree of freedom of the QCD gauge field, and thus present an
attractive candidate for study.  In $\mathrm{SU}(N)$ gauge theory, centre
vortices have a clear theoretical link to confinement
\cite{Greensite:2007zz,'tHooft:1977hy,'tHooft:1979uj}, and in $\mathrm{SU}(2)$
they have been shown to be responsible for dynamical chiral symmetry
breaking \cite{Bowman:2008qd,Hoellwieser:2014isa,Hollwieser:2013xja,deForcrand:1999ms,Engelhardt:2002qs,Bornyakov:2007fz,Alexandrou:1999vx,Kovalenko:2005rz} through lattice-QCD simulations.

In $\mathrm{SU}(3)$ gauge theory, the picture is less clear.  While lattice
results have shown a loss of string tension, and thus confinement, to be
coincident with the removal of centre vortices, 
a background consisting solely of centre vortices was shown to reproduce just $67\%$ of the string tension \cite{Langfeld:2003ev}.%

While there has been extensive work within $\mathrm{SU}(2),$ lattice studies of a connection between dynamical chiral symmetry breaking and centre vortices in $\mathrm{SU}(3)$ \cite{Ilgenfritz:2007ua,Bowman:2010zr,OMalley:2011aa} are few in number. It has been shown that topological charge density in $\mathrm{SU}(3)$ gauge theory lies preferentially near centre
vortices \cite{Ilgenfritz:2007ua}, thus suggesting a similar role for centre vortices to that observed in $\mathrm{SU}(2)$.

Study of dynamical mass generation through the quark propagator in $\mathrm{SU}(3)$ \cite{Bowman:2010zr} appeared to show the persistence of dynamical mass generation after vortex removal. The use of the AsqTad staggered fermion action, however, which explicitly breaks chiral symmetry, leads to a corresponding insensitivity to dynamical chiral symmetry breaking effects.

In Ref.~\cite{OMalley:2011aa} the ground-state hadron spectrum was studied using a Wilson fermion action. There, vortex removal resulted in an absence of dynamical chiral symmetry breaking, and degeneracy of the ground state meson and baryon spectra. The Wilson fermion action, again, explicitly breaks chiral symmetry on the lattice, and so results in additive mass renormalisation, and thus an unknown lattice bare quark mass.

Here we study the quark propagator in $\mathrm{SU}(3)$ using the
overlap-Dirac fermion action, which possesses an exact
(lattice-deformed) chiral symmetry. Using a fermion action that
respects chiral symmetry on the lattice allows us to examine dynamical
chiral symmetry breaking in $\mathrm{SU}(3)$ gauge theory for the
first time.

The Landau-gauge quark propagator has often
\cite{Bowman:2005zi,Roberts:2007ji,Alkofer:2000wg,Fischer:2006ub} been used as a probe of dynamical
chiral symmetry breaking.  At low momenta, enhancement of the Dirac
scalar part of the propagator, commonly referred to as the mass
function and associated with the concept of a constituent quark mass,
provides a clear signal of the presence or absence of dynamical chiral
symmetry breaking.
In $\mathrm{SU}(2)$ gauge theory, the mass function clearly displays the absence
of dynamical chiral symmetry breaking upon centre vortex removal, as
the mass function does not develop a dynamically generated mass in the
infrared limit \cite{Bowman:2008qd}.  However, a similar study in $\mathrm{SU}(3)$ gauge theory using the
AsqTad-quark \cite{Orginos:1999cr}  propagator did not reveal a comparable role in dynamical chiral symmetry breaking in $\mathrm{SU}(3)$ \cite{Bowman:2010zr}.  There the mass function sustained
dynamical mass generation on vortex-removed configurations.  However
in Ref.~\cite{OMalley:2011aa}, where the vortex-removed hadron
spectrum was studied with Wilson fermions, it became clear that this residual mass
generation on vortex-removed configurations was not associated with
chiral symmetry; {\it i.e.} that chiral symmetry was indeed lost upon
vortex removal.

Both the AsqTad- and Wilson-fermion actions explicitly break chiral
symmetry, and hence the relation between centre vortices and dynamical chiral
symmetry breaking may be clouded by the resulting lattice artefacts.  In this
letter we study the Landau gauge quark propagator using the superior
chiral properties of the overlap-Dirac fermion action.  The overlap
fermion action provides a realisation of chiral symmetry on the
lattice and is renowned for its sensitivity to the topological
structure of the gauge fields. We find for the first time a loss of dynamical mass generation in the
Landau-gauge quark propagator coincident with vortex removal in $\mathrm{SU}(3)$ gauge theory.  Through
a study of the topological charge density under ${\cal
  O}(a^4)$-improved cooling, we are able to trace this success to the
overlap-fermion's sensitivity to the subtle manner in which instanton
degrees of freedom are compromised through the process of centre
vortex removal.

We also demonstrate how the centre vortex degrees of freedom can
reproduce dynamical mass generation after smoothing the vortex-only
gauge fields with improved cooling.  We observe dynamical mass
generation on the vortex-only backgrounds consistent with that on the
original gauge-field ensemble following the same amount of smoothing.

Through visualizations of the instanton-like degrees of freedom in the
various gauge-field ensembles, we find evidence of a link
between the centre vortex and instanton structure of the vacuum.
While vortex removal destabilizes instanton-like objects under ${\cal
  O}(a^4)$-improved cooling, vortex-only backgrounds provide
gauge-field degrees of freedom sufficient to create instantons upon
cooling.

\section{Centre Vortices on the Lattice}

We study centre vortices on the lattice in the standard way,
commencing with gauge transformations designed to bring the lattice
link variables,
\begin{equation}
U_{\mu}(x) = {\cal P}\, \exp\left ( ig \int_0^a d\lambda\, A_\mu(x + \lambda
\hat\mu)\, \right ) \, ,
\end{equation}
to be as close as possible to centre elements of $\mathrm{SU}(3)$ via Maximal Centre Gauge (MCG)
\cite{Montero:1999by},
%
%
then projecting onto the $\mathbb{Z}_3$ centre-subgroup \cite{Montero:1999by,DelDebbio:1998uu,DelDebbio:1996mh,Vink:1992ys,Alexandrou:1999vx,Faber:2001zs} to produce the vortex-only configuration,
\begin{equation}
Z_{\mu}(x) = \exp{\bigg [}\frac{2\pi i}{3}\, m_{\mu}(x){\bigg
]}\, \mathrm{I}\, \mathrm{,} \quad m_{\mu} \in \{-1,0,1\} \, .
\end{equation}


The vortices are identified by the centre charge, $z$, found
by taking the product of the links around a plaquette,
\begin{equation}
z = \frac{1}{3} \mbox{Tr} \prod_\Box Z_\mu(x) = \exp \left ( 2 \pi i\, \frac{n}{3} \right ) \, .
\label{CentreCharge}
\end{equation}
If $z=1$, no vortex pierces the plaquette. If $z \neq 1$, a vortex with
charge $z$ pierces the plaquette.  In the smooth gauge-field limit,
all links approach the identity, and no vortices are found.
Vortices are identified as the defects in the centre-projected gauge
field.

Upon transforming each link to the closest element of the centre
$Z_{\mu}(x)$, we are able to define three ensembles:
\begin{enumerate}
\item The original `untouched' configurations, $U_{\mu}(x),$

\item The projected vortex-only configurations, $Z_{\mu}(x),$

\item The vortex-removed configurations, $Z^{\dagger}_{\mu}(x)\,
  U_{\mu}(x).$
\end{enumerate}
Each of the ensembles are gauge-fixed to Landau gauge. 
By comparing results on these three ensembles, we are able to isolate
the effects of centre vortices.

\section{Landau-Gauge Overlap Quark Propagator}
We calculate the quark propagator using the overlap fermion operator,
which satisfies the Ginsparg-Wilson relation \cite{Ginsparg:1981bj},
and thus provides a lattice-realisation of chiral symmetry.  It has a
superior sensitivity to gauge field topology than the aforementioned
AsqTad and Wilson lattice fermion operators.  Explicitly, the massive
overlap-Dirac operator \cite{Narayanan:1993ss,Narayanan:1994gw} is
given by
\begin{equation}
\label{massov}
D_{o}(\mu) = \frac{(1-\mu)}{2}\, \left [ 1 + \gamma_{5}\,
  \epsilon \left ( \gamma_{5}\, D(-m_{w}) \right ) \right ] + \mu\, ,
\end{equation}
where $\epsilon$ is the matrix sign function.  We use the fat-link
irrelevant clover (FLIC) fermion operator
\cite{Zanotti:2001yb,Kamleh:2004aw,Kamleh:2004xk,Kamleh:2001ff} as
the overlap kernel $D(-m_{w})$, with regulator parameter $m_{w} = 1$.
The overlap mass parameter, $\mu = 0.004$, provides a bare quark mass
of $12$ MeV for our calculations.

In a covariant gauge, the lattice quark propagator can be decomposed
into Dirac scalar and vector components as
\begin{equation}
S(p) = \frac{Z(p)}{iq\!\!\!/\ + M(p)}\, ,
\end{equation}
with $M(p)$ the non-perturbative mass function and $Z(p)$ containing
all renormalisation information.  The infrared behaviour of $M(p)$
reveals the presence or absence of dynamical mass generation, and thus
of dynamical chiral symmetry breaking. 


To isolate the role of centre symmetry, results are calculated on $50$
pure gauge-field configurations using the L{\"u}scher-Weisz
$\mathcal{O}(a^{2})$ mean-field improved action \cite{Luscher:1984xn},
with a $20^{3} \times 40$ volume at a lattice spacing of $0.125 \,
\mathrm{fm}$. We fix to Landau gauge using a Fourier transform
accelerated algorithm \cite{Davies:1987vs}, fixing to the
$\mathcal{O}(a^2)$ improved gauge-fixing functional
\cite{Bonnet:1999mj}. The vortex-only configurations are
pre-conditioned with a random gauge transformation before gauge-fixing
for improved algorithmic convergence. A cylinder cut
\cite{Leinweber:1998im} is performed on propagator data, and $Z(p)$ is
renormalised to be $1$ at the highest momentum considered, $p \simeq
5.2$ GeV.

\begin{figure*}[thb]
\subfigure[]{
\includegraphics[width=\columnwidth]{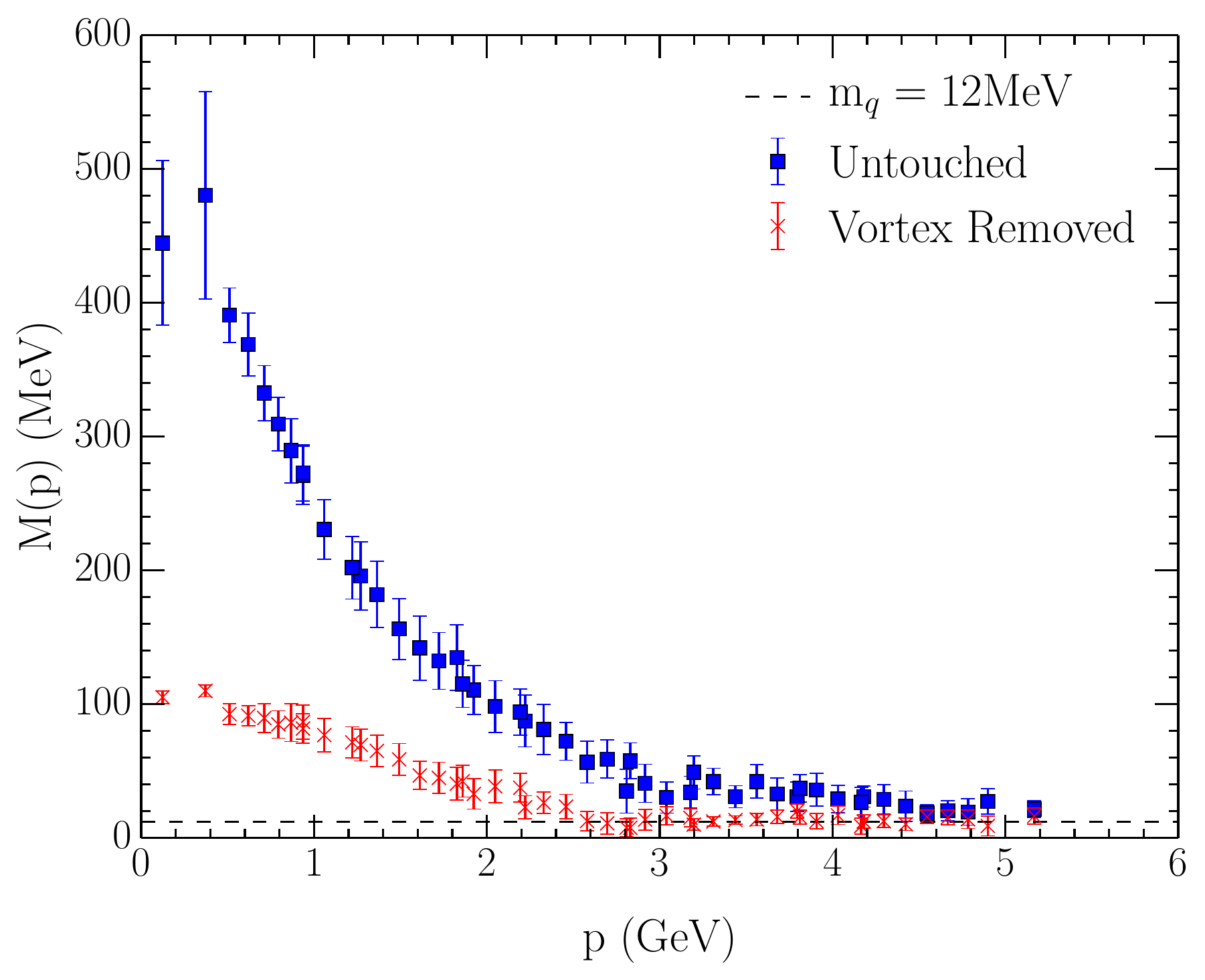}}
\subfigure[]{
\includegraphics[width=\columnwidth]{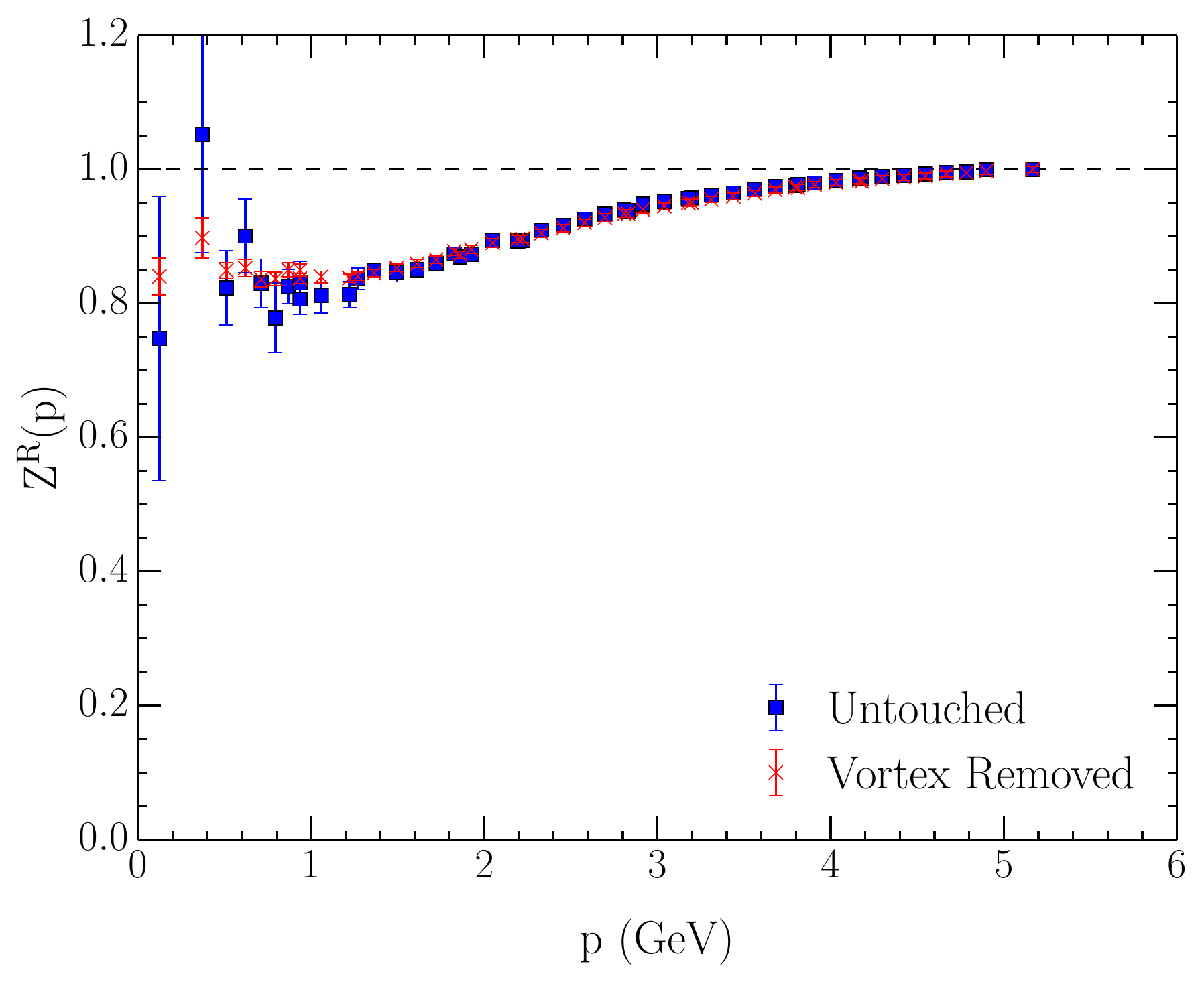}}
\caption{The mass (a) and renormalisation (b) functions on the
  original (untouched) (squares) and vortex-removed (crosses)
  configurations.  Removal of the vortex structure from the gauge
  fields spoils dynamical mass generation and thus dynamical chiral
  symmetry breaking.}
\label{M00400UTVR}
\end{figure*}

Results for the untouched and vortex-removed ensembles are plotted in
Fig.~\ref{M00400UTVR}.  The renormalisation function shows similar
behaviour in both the untouched and vortex-removed cases.  However,
the mass function reveals a significant change upon vortex removal.

On the untouched ensemble, the mass function shows strong enhancement
in the infrared, displaying the presence of dynamical mass generation.
By contrast, dynamical mass generation is largely
suppressed upon vortex removal with only a relatively small level of residual infrared
enhancement remaining \footnote{Our studies of the topological charge
  density of the vortex-removed configurations suggest that this residual
  enhancement in the mass function is likely associated with
  imperfections in the identification of all centre vortices in the
  MCG vortex-removal procedure.}.  Unlike the AsqTad propagator, which
showed little to no change in the infrared enhancement
\cite{Bowman:2010zr}, the overlap operator is able to `see' the subtle
damage caused to the gauge fields through vortex removal. The removal
of the vortex structure from gauge fields has spoiled dynamical mass
generation, and thus dynamical chiral symmetry breaking.


The smoothness requirement of the overlap operator
\cite{Narayanan:1994gw} contrasts the rough nature of vortex-only
configurations consisting solely of centre elements, and the
overlap fermion action is thus not well defined on vortex-only
configurations.  To address this issue we smooth the gauge-field
configurations. This is additionally motivated by evidence that, in
$\mathrm{SU}(2)$ gauge theory, vortex-only configurations are too rough to
reproduce the low-lying modes of the Dirac operator essential to
dynamical chiral symmetry breaking, but are able to do so after
smearing \cite{Hollwieser:2008tq}.  Smoothing is performed using
three-loop $\mathcal{O}(a^4)$-improved cooling
\cite{BilsonThompson:2002jk}.  

\begin{figure}[thb]
\includegraphics[width=\columnwidth]{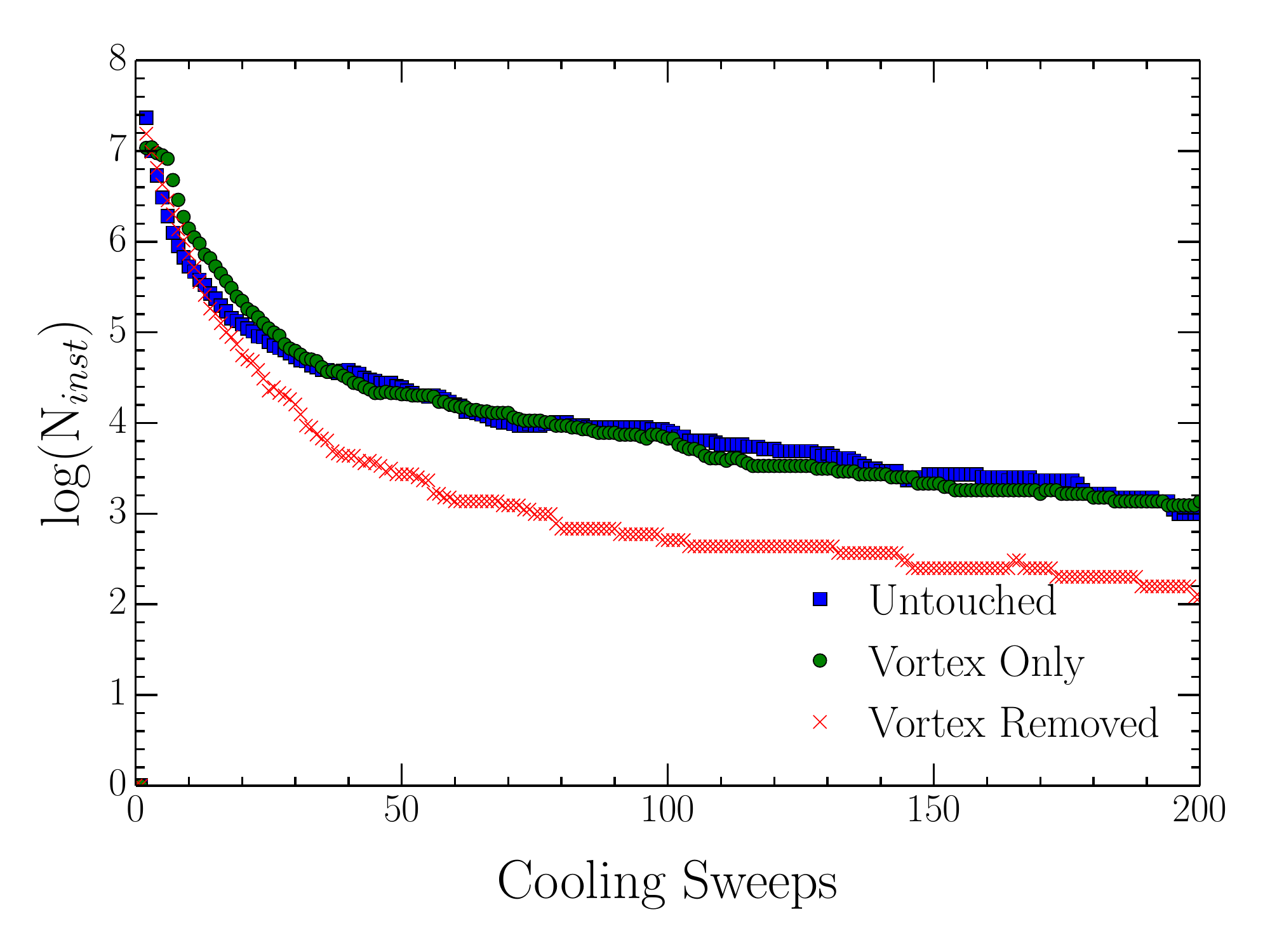}
\caption{A $\log$ plot of the number of instanton-like objects per
  configuration found on untouched, vortex-only and vortex-removed
  ensembles as a function of $\mathcal{O}(a^4)$-improved cooling
  sweeps.}
\label{num}
\end{figure}

By examining the local maxima of the action density on vortex-only
configurations during cooling, we find that after just 10 sweeps of
smoothing these local maxima stabilise and begin to resemble classical
instantons in shape and corresponding topological charge density at
the centre \cite{Moran:2008qd}.
The average number of these maxima per configuration is plotted in
Fig.~\ref{num} as a proxy for the number of instanton-like objects per
configuration for up to 200 sweeps.  The number of objects found on
untouched and vortex-only configurations remains very similar even
after large amounts of cooling.  

In contrast, the number of objects on vortex-removed configurations is
greatly reduced.  Vortex-removal has destabilised the otherwise
topologically-nontrivial instanton-like objects.  Early in the
smoothing procedure the topological charge density of the vortex
removed configurations qualitatively resembles that of the original
configurations. It is perhaps unsurprising that a fermion operator
that is not sensitive to the spoiling of instanton-like objects
through vortex-removal would erroneously report little change to
dynamical mass generation.  It is remarkable that the overlap operator
is sensitive to the subtle changes of vortex removal in the absence of
any smoothing.

\begin{figure*}[thb]
\subfigure[]{
\includegraphics[width=\columnwidth]{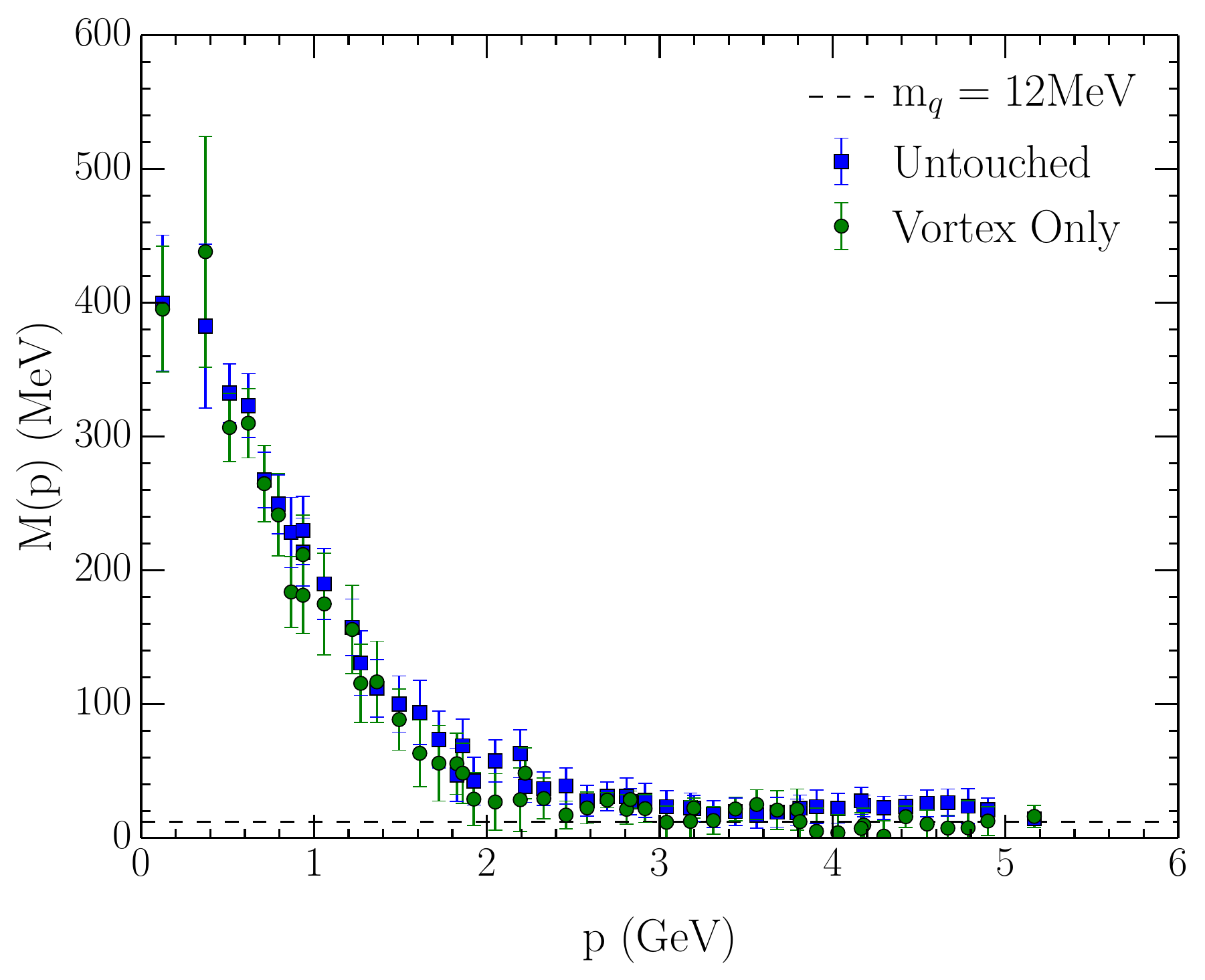}}
\subfigure[]{
\includegraphics[width=\columnwidth]{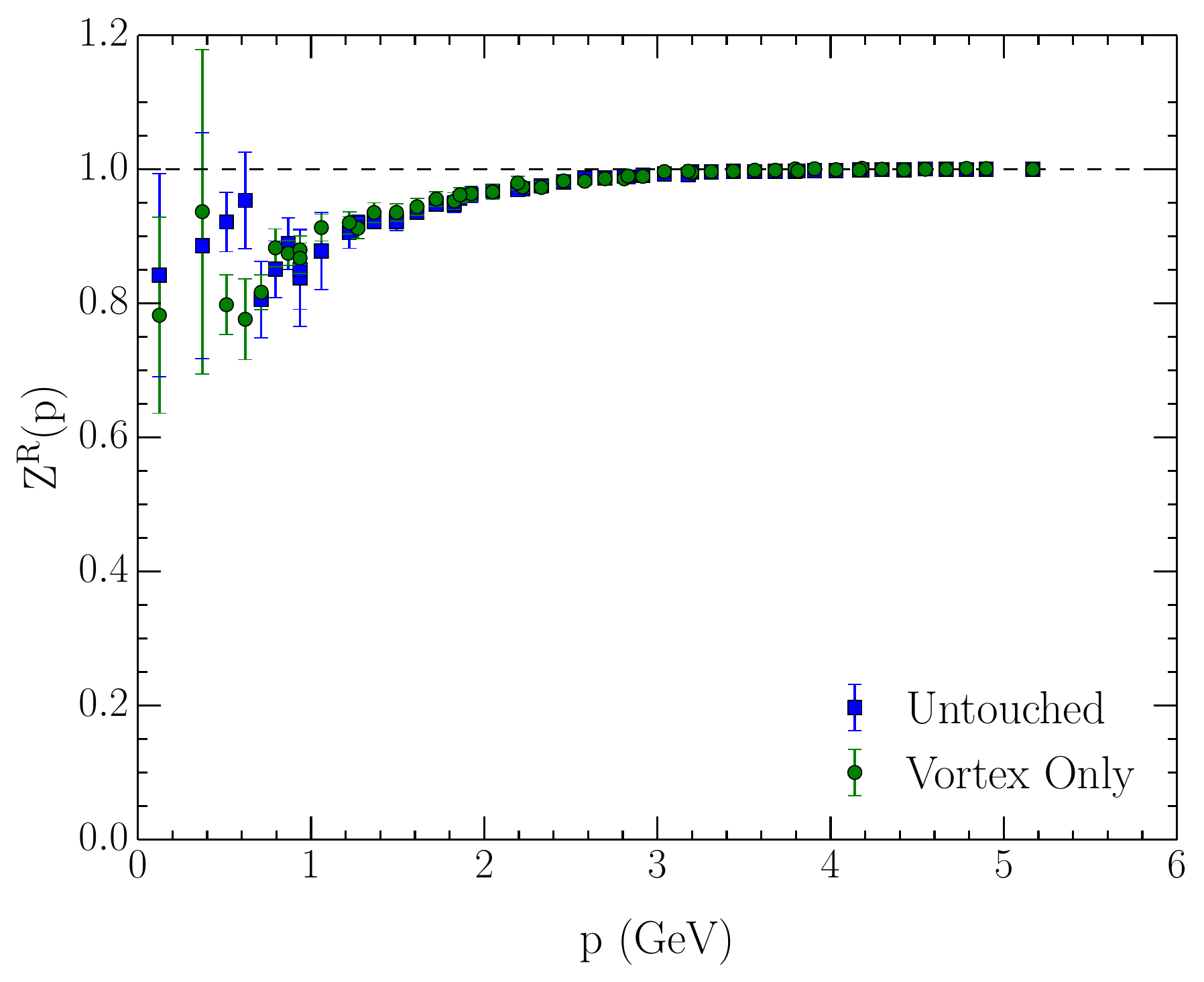}}
\caption{The mass(a) and renormalisation (b) functions on the original (untouched) (squares)
  and vortex-only (circles) configurations after 10 sweeps of three-loop
  $\mathcal{O}(a^4)$-improved cooling, at an input bare quark mass of
  $12$ MeV.}
\label{M00400UTcVOc}
\end{figure*}

Although there does not appear to be a one-to-one connection between the
backgrounds dominated by instanton-like objects found in the untouched
and vortex-only cases on a configuration-by-configuration basis, the
objects are qualitatively similar in number and size. 
Despite consisting solely of the centre elements, the centre vortex
information encapsulates the qualitative essence of the QCD vacuum
structure.  It contains the `seeds' of instantons, which are
reproduced through cooling.

Just as the centre-vortex information alone was sufficient to
reproduce instantons through cooling, vortex removal is sufficient to
destroy the stability of instantons under cooling, with the vast
majority of topological objects being removed as seen in Fig.~\ref{num}.
On vortex-removed backgrounds a few instanton-like objects remain,
which closely resemble those found on the untouched and vortex-only
backgrounds in size despite their greatly reduced number. These
residual objects provide a mechanism for the remnant of
dynamical mass generation seen on vortex-removed configurations in
Fig.~\ref{M00400UTVR}.

To quantify the extent to which the centre-vortex information
encapsulated in the vortex-only configurations can give rise to
dynamical mass generation, we calculate the overlap quark propagator
on both untouched and vortex-only configurations after 10 sweeps of
cooling.  The mass and renormalisation functions are illustrated in
Fig.~\ref{M00400UTcVOc}.  As expected, the mass function on the
untouched configurations shows some reduction in dynamical mass
generation, while being qualitatively similar to the uncooled results
\cite{Trewartha:2013qga}.  The vortex-only results for the mass
function are strikingly similar to the untouched; they show the vortex-only configurations reproducing almost all dynamical mass generation.  The renormalisation functions
also share a similar behavior.  The background of instanton-like
objects emerging from the vortex-only configurations under cooling is
able to reproduce the features of the quark propagator on the full
backgrounds.
\section{Conclusion}

Combined, these results provide evidence that the centre vortex structure of the vacuum plays a fundamental role in dynamical chiral symmetry breaking in
$\mathrm{SU}(3)$ gauge theory.   For the first time, we have demonstrated the
removal of dynamical mass generation via the removal of the
centre-vortex degrees of freedom from the gauge fields.  Moreover, we
have demonstrated how the vortex-only degrees of freedom encapsulate
the qualitative features of the gauge fields, reproducing the average
number and size of instanton-like objects under smoothing via
cooling.  These features reproduce the dynamical mass generation
observed on the original gauge fields following the same smoothing.

We have also found a link between the stability of instanton-like
objects under cooling and centre vortex removal.  Vortex removal
spoils and destabilizes instantons, resulting in them being quickly
removed from the lattice under cooling.  Correspondingly, vortex-only
configurations quickly produce a background of instanton-like objects
with general features resembling those found in the untouched case.
Our results are consistent with the instanton model of dynamical
mass generation, and illustrate a connection between the
centre-vortex structure and the instanton structure of $\mathrm{SU}(3)$ gauge
fields.  Our findings are similar to a connection shown in $\mathrm{SU}(2)$
gauge theory \cite{Hollwieser:2014osa}.  In conclusion, these results support the hypothesis that centre
vortices are the fundamental long-range structures underpinning dynamical
chiral symmetry breaking in $\mathrm{SU}(3)$ gauge theory.

\section{Acknowledgements}
This research was undertaken with the
assistance of resources awarded at the NCI National Facility in
Canberra, Australia, and the iVEC facilities at Murdoch University
(iVEC@Murdoch) and the University of Western Australia
(iVEC@UWA). These resources are provided through the National
Computational Merit Allocation Scheme and the University of Adelaide
Partner Share supported by the Australian Government.  We also
acknowledge eResearch SA for their support of our supercomputers.
This research is supported by the Australian Research Council through
grants DP120104627, DP150103164, LE120100181 and LE110100234.

\bibliography{PLB020415}

\begin{thebibliography}{10}
\expandafter\ifx\csname url\endcsname\relax
  \def\url#1{\texttt{#1}}\fi
\expandafter\ifx\csname urlprefix\endcsname\relax\def\urlprefix{URL }\fi
\expandafter\ifx\csname href\endcsname\relax
  \def\href#1#2{#2} \def\path#1{#1}\fi

\bibitem{Belavin:1975fg}
A.~Belavin, A.~M. Polyakov, A.~Schwartz, Y.~Tyupkin, {Pseudoparticle Solutions
  of the Yang-Mills Equations}, Phys.Lett. B59 (1975) 85--87.
\newblock \href {http://dx.doi.org/10.1016/0370-2693(75)90163-X}
  {\path{doi:10.1016/0370-2693(75)90163-X}}.

\bibitem{Polyakov:1976fu}
A.~M. Polyakov, {Quark Confinement and Topology of Gauge Groups}, Nucl.Phys.
  B120 (1977) 429--458.
\newblock \href {http://dx.doi.org/10.1016/0550-3213(77)90086-4}
  {\path{doi:10.1016/0550-3213(77)90086-4}}.

\bibitem{Mandelstam:1974pi}
S.~Mandelstam, {Vortices and Quark Confinement in Nonabelian Gauge Theories},
  Phys.Rept. 23 (1976) 245--249.
\newblock \href {http://dx.doi.org/10.1016/0370-1573(76)90043-0}
  {\path{doi:10.1016/0370-1573(76)90043-0}}.

\bibitem{Jackiw:1976pf}
R.~Jackiw, C.~Rebbi, {Vacuum Periodicity in a Yang-Mills Quantum Theory},
  Phys.Rev.Lett. 37 (1976) 172--175.
\newblock \href {http://dx.doi.org/10.1103/PhysRevLett.37.172}
  {\path{doi:10.1103/PhysRevLett.37.172}}.

\bibitem{Callan:1976je}
J.~Callan, Curtis~G., R.~Dashen, D.~J. Gross, {The Structure of the Gauge
  Theory Vacuum}, Phys.Lett. B63 (1976) 334--340.
\newblock \href {http://dx.doi.org/10.1016/0370-2693(76)90277-X}
  {\path{doi:10.1016/0370-2693(76)90277-X}}.

\bibitem{Schafer:1996wv}
T.~Sch{\"a}fer, E.~V. Shuryak, {Instantons in QCD}, Rev.Mod.Phys. 70 (1998)
  323--426.
\newblock \href {http://arxiv.org/abs/hep-ph/9610451}
  {\path{arXiv:hep-ph/9610451}}, \href
  {http://dx.doi.org/10.1103/RevModPhys.70.323}
  {\path{doi:10.1103/RevModPhys.70.323}}.

\bibitem{Trewartha:2013qga}
D.~Trewartha, W.~Kamleh, D.~Leinweber, D.~S. Roberts, {Quark Propagation in the
  Instantons of Lattice QCD}, Phys.Rev. D88 (2013) 034501.
\newblock \href {http://arxiv.org/abs/1306.3283} {\path{arXiv:1306.3283}},
  \href {http://dx.doi.org/10.1103/PhysRevD.88.034501}
  {\path{doi:10.1103/PhysRevD.88.034501}}.

\bibitem{'tHooft:1977hy}
G.~'t~Hooft, {On the Phase Transition Towards Permanent Quark Confinement},
  Nucl.Phys. B138 (1978) 1.
\newblock \href {http://dx.doi.org/10.1016/0550-3213(78)90153-0}
  {\path{doi:10.1016/0550-3213(78)90153-0}}.

\bibitem{'tHooft:1979uj}
G.~'t~Hooft, {A Property of Electric and Magnetic Flux in Nonabelian Gauge
  Theories}, Nucl.Phys. B153 (1979) 141.
\newblock \href {http://dx.doi.org/10.1016/0550-3213(79)90595-9}
  {\path{doi:10.1016/0550-3213(79)90595-9}}.

\bibitem{Cornwall:1979hz}
J.~M. Cornwall, {Quark Confinement and Vortices in Massive Gauge Invariant
  QCD}, Nucl.Phys. B157 (1979) 392.
\newblock \href {http://dx.doi.org/10.1016/0550-3213(79)90111-1}
  {\path{doi:10.1016/0550-3213(79)90111-1}}.

\bibitem{Nielsen:1979xu}
H.~B. Nielsen, P.~Olesen, {A Quantum Liquid Model for the QCD Vacuum: Gauge and
  Rotational Invariance of Domained and Quantized Homogeneous Color Fields},
  Nucl.Phys. B160 (1979) 380.
\newblock \href {http://dx.doi.org/10.1016/0550-3213(79)90065-8}
  {\path{doi:10.1016/0550-3213(79)90065-8}}.

\bibitem{Ambjorn:1980ms}
J.~Ambjorn, P.~Olesen, {A Color Magnetic Vortex Condensate in QCD}, Nucl.Phys.
  B170 (1980) 265.
\newblock \href {http://dx.doi.org/10.1016/0550-3213(80)90150-9}
  {\path{doi:10.1016/0550-3213(80)90150-9}}.

\bibitem{Greensite:2014gra}
J.~Greensite, R.~H{\"o}llwieser, {Double-winding Wilson loops and monopole
  confinement mechanisms}\href {http://arxiv.org/abs/1411.5091}
  {\path{arXiv:1411.5091}}.

\bibitem{Greensite:2007zz}
J.~Greensite, {Center vortices, and other scenarios of quark confinement},
  Eur.Phys.J.ST 140 (2007) 1--52.
\newblock \href {http://dx.doi.org/10.1140/epjst/e2007-00002-6}
  {\path{doi:10.1140/epjst/e2007-00002-6}}.

\bibitem{Bowman:2008qd}
P.~O. Bowman, K.~Langfeld, D.~B. Leinweber, A.~O'~Cais, A.~Sternbeck, et~al.,
  {Center vortices and the quark propagator in SU(2) gauge theory}, Phys.Rev.
  D78 (2008) 054509.
\newblock \href {http://arxiv.org/abs/0806.4219} {\path{arXiv:0806.4219}},
  \href {http://dx.doi.org/10.1103/PhysRevD.78.054509}
  {\path{doi:10.1103/PhysRevD.78.054509}}.

\bibitem{Hoellwieser:2014isa}
R.~H{\"o}llwieser, M.~Faber, T.~Schweigler, U.~M. Heller, {Chiral Symmetry
  Breaking from Center Vortices}, PoS LATTICE2013 (2014) 505.

\bibitem{Hollwieser:2013xja}
R.~H{\"o}llwieser, T.~Schweigler, M.~Faber, U.~M. Heller, {Center Vortices and
  Chiral Symmetry Breaking in SU(2) Lattice Gauge Theory}, Phys.Rev. D88 (2013)
  114505.
\newblock \href {http://arxiv.org/abs/1304.1277} {\path{arXiv:1304.1277}},
  \href {http://dx.doi.org/10.1103/PhysRevD.88.114505}
  {\path{doi:10.1103/PhysRevD.88.114505}}.

\bibitem{deForcrand:1999ms}
P.~de~Forcrand, M.~D'Elia, {On the relevance of center vortices to QCD},
  Phys.Rev.Lett. 82 (1999) 4582--4585.
\newblock \href {http://arxiv.org/abs/hep-lat/9901020}
  {\path{arXiv:hep-lat/9901020}}, \href
  {http://dx.doi.org/10.1103/PhysRevLett.82.4582}
  {\path{doi:10.1103/PhysRevLett.82.4582}}.

\bibitem{Engelhardt:2002qs}
M.~Engelhardt, {Center vortex model for the infrared sector of Yang-Mills
  theory: Quenched Dirac spectrum and chiral condensate}, Nucl.Phys. B638
  (2002) 81--110.
\newblock \href {http://arxiv.org/abs/hep-lat/0204002}
  {\path{arXiv:hep-lat/0204002}}, \href
  {http://dx.doi.org/10.1016/S0550-3213(02)00470-4}
  {\path{doi:10.1016/S0550-3213(02)00470-4}}.

\bibitem{Bornyakov:2007fz}
V.~Bornyakov, E.-M. Ilgenfritz, B.~Martemyanov, S.~Morozov,
  M.~Muller-Preussker, et~al., {Interrelation between monopoles, vortices,
  topological charge and chiral symmetry breaking: Analysis using overlap
  fermions for SU(2)}, Phys.Rev. D77 (2008) 074507.
\newblock \href {http://arxiv.org/abs/0708.3335} {\path{arXiv:0708.3335}},
  \href {http://dx.doi.org/10.1103/PhysRevD.77.074507}
  {\path{doi:10.1103/PhysRevD.77.074507}}.

\bibitem{Alexandrou:1999vx}
C.~Alexandrou, P.~de~Forcrand, M.~D'Elia, {The Role of center vortices in QCD},
  Nucl.Phys. A663 (2000) 1031--1034.
\newblock \href {http://arxiv.org/abs/hep-lat/9909005}
  {\path{arXiv:hep-lat/9909005}}, \href
  {http://dx.doi.org/10.1016/S0375-9474(99)00763-0}
  {\path{doi:10.1016/S0375-9474(99)00763-0}}.

\bibitem{Kovalenko:2005rz}
A.~Kovalenko, S.~Morozov, M.~Polikarpov, V.~Zakharov, {On topological
  properties of vacuum defects in lattice Yang-Mills theories}, Phys.Lett. B648
  (2007) 383--387.
\newblock \href {http://arxiv.org/abs/hep-lat/0512036}
  {\path{arXiv:hep-lat/0512036}}, \href
  {http://dx.doi.org/10.1016/j.physletb.2007.03.036}
  {\path{doi:10.1016/j.physletb.2007.03.036}}.

\bibitem{Langfeld:2003ev}
K.~Langfeld, {Vortex structures in pure SU(3) lattice gauge theory}, Phys.Rev.
  D69 (2004) 014503.
\newblock \href {http://arxiv.org/abs/hep-lat/0307030}
  {\path{arXiv:hep-lat/0307030}}, \href
  {http://dx.doi.org/10.1103/PhysRevD.69.014503}
  {\path{doi:10.1103/PhysRevD.69.014503}}.

\bibitem{Ilgenfritz:2007ua}
E.-M. Ilgenfritz, K.~Koller, Y.~Koma, G.~Schierholz, T.~Streuer, et~al.,
  {Localization of overlap modes and topological charge, vortices and monopoles
  in SU(3) LGT}, PoS LAT2007 (2007) 311.
\newblock \href {http://arxiv.org/abs/0710.2607} {\path{arXiv:0710.2607}}.

\bibitem{Bowman:2010zr}
P.~O. Bowman, K.~Langfeld, D.~B. Leinweber, A.~Sternbeck, L.~von Smekal,
  et~al., {Role of center vortices in chiral symmetry breaking in SU(3) gauge
  theory}, Phys.Rev. D84 (2011) 034501.
\newblock \href {http://arxiv.org/abs/1010.4624} {\path{arXiv:1010.4624}},
  \href {http://dx.doi.org/10.1103/PhysRevD.84.034501}
  {\path{doi:10.1103/PhysRevD.84.034501}}.

\bibitem{OMalley:2011aa}
E.-A. O'Malley, W.~Kamleh, D.~Leinweber, P.~Moran, {SU(3) centre vortices
  underpin confinement and dynamical chiral symmetry breaking}, Phys.Rev. D86
  (2012) 054503.
\newblock \href {http://arxiv.org/abs/1112.2490} {\path{arXiv:1112.2490}},
  \href {http://dx.doi.org/10.1103/PhysRevD.86.054503}
  {\path{doi:10.1103/PhysRevD.86.054503}}.

\bibitem{Bowman:2005zi}
P.~Bowman, U.~Heller, D.~Leinweber, A.~Williams, J.~Zhang, {Quark propagator
  from LQCD and its physical implications}, Lect.Notes Phys. 663 (2005) 17--63.
\newblock \href {http://dx.doi.org/10.1007/11356462_2}
  {\path{doi:10.1007/11356462_2}}.

\bibitem{Roberts:2007ji}
C.~Roberts, {Hadron Properties and Dyson-Schwinger Equations},
  Prog.Part.Nucl.Phys. 61 (2008) 50--65.
\newblock \href {http://arxiv.org/abs/0712.0633} {\path{arXiv:0712.0633}},
  \href {http://dx.doi.org/10.1016/j.ppnp.2007.12.034}
  {\path{doi:10.1016/j.ppnp.2007.12.034}}.

\bibitem{Alkofer:2000wg}
R.~Alkofer, L.~von Smekal, {The Infrared behavior of QCD Green's functions:
  Confinement dynamical symmetry breaking, and hadrons as relativistic bound
  states}, Phys.Rept. 353 (2001) 281.
\newblock \href {http://arxiv.org/abs/hep-ph/0007355}
  {\path{arXiv:hep-ph/0007355}}, \href
  {http://dx.doi.org/10.1016/S0370-1573(01)00010-2}
  {\path{doi:10.1016/S0370-1573(01)00010-2}}.

\bibitem{Fischer:2006ub}
C.~S. Fischer, {Infrared properties of QCD from Dyson-Schwinger equations},
  J.Phys. G32 (2006) R253--R291.
\newblock \href {http://arxiv.org/abs/hep-ph/0605173}
  {\path{arXiv:hep-ph/0605173}}, \href
  {http://dx.doi.org/10.1088/0954-3899/32/8/R02}
  {\path{doi:10.1088/0954-3899/32/8/R02}}.

\bibitem{Orginos:1999cr}
K.~Orginos, D.~Toussaint, R.~Sugar, {Variants of fattening and flavor symmetry
  restoration}, Phys.Rev. D60 (1999) 054503.
\newblock \href {http://arxiv.org/abs/hep-lat/9903032}
  {\path{arXiv:hep-lat/9903032}}, \href
  {http://dx.doi.org/10.1103/PhysRevD.60.054503}
  {\path{doi:10.1103/PhysRevD.60.054503}}.

\bibitem{Montero:1999by}
A.~Montero, {Study of SU(3) vortex - like configurations with a new maximal
  center gauge fixing method}, Phys.Lett. B467 (1999) 106--111.
\newblock \href {http://arxiv.org/abs/hep-lat/9906010}
  {\path{arXiv:hep-lat/9906010}}, \href
  {http://dx.doi.org/10.1016/S0370-2693(99)01113-2}
  {\path{doi:10.1016/S0370-2693(99)01113-2}}.

\bibitem{DelDebbio:1998uu}
L.~Del~Debbio, M.~Faber, J.~Giedt, J.~Greensite, S.~Olejnik, {Detection of
  center vortices in the lattice Yang-Mills vacuum}, Phys.Rev. D58 (1998)
  094501.
\newblock \href {http://arxiv.org/abs/hep-lat/9801027}
  {\path{arXiv:hep-lat/9801027}}, \href
  {http://dx.doi.org/10.1103/PhysRevD.58.094501}
  {\path{doi:10.1103/PhysRevD.58.094501}}.

\bibitem{DelDebbio:1996mh}
L.~Del~Debbio, M.~Faber, J.~Greensite, S.~Olejnik, {Center dominance and Z(2)
  vortices in SU(2) lattice gauge theory}, Phys.Rev. D55 (1997) 2298--2306.
\newblock \href {http://arxiv.org/abs/hep-lat/9610005}
  {\path{arXiv:hep-lat/9610005}}, \href
  {http://dx.doi.org/10.1103/PhysRevD.55.2298}
  {\path{doi:10.1103/PhysRevD.55.2298}}.

\bibitem{Vink:1992ys}
J.~C. Vink, U.-J. Wiese, {Gauge fixing on the lattice without ambiguity},
  Phys.Lett. B289 (1992) 122--126.
\newblock \href {http://arxiv.org/abs/hep-lat/9206006}
  {\path{arXiv:hep-lat/9206006}}, \href
  {http://dx.doi.org/10.1016/0370-2693(92)91372-G}
  {\path{doi:10.1016/0370-2693(92)91372-G}}.

\bibitem{Faber:2001zs}
M.~Faber, J.~Greensite, S.~Olejnik, {Direct Laplacian center gauge}, JHEP 0111
  (2001) 053.
\newblock \href {http://arxiv.org/abs/hep-lat/0106017}
  {\path{arXiv:hep-lat/0106017}}, \href
  {http://dx.doi.org/10.1088/1126-6708/2001/11/053}
  {\path{doi:10.1088/1126-6708/2001/11/053}}.

\bibitem{Ginsparg:1981bj}
P.~H. Ginsparg, K.~G. Wilson, {A Remnant of Chiral Symmetry on the Lattice},
  Phys.Rev. D25 (1982) 2649.
\newblock \href {http://dx.doi.org/10.1103/PhysRevD.25.2649}
  {\path{doi:10.1103/PhysRevD.25.2649}}.

\bibitem{Narayanan:1993ss}
R.~Narayanan, H.~Neuberger, {Chiral fermions on the lattice}, Phys.Rev.Lett. 71
  (1993) 3251--3254.
\newblock \href {http://arxiv.org/abs/hep-lat/9308011}
  {\path{arXiv:hep-lat/9308011}}, \href
  {http://dx.doi.org/10.1103/PhysRevLett.71.3251}
  {\path{doi:10.1103/PhysRevLett.71.3251}}.

\bibitem{Narayanan:1994gw}
R.~Narayanan, H.~Neuberger, {A Construction of lattice chiral gauge theories},
  Nucl.Phys. B443 (1995) 305--385.
\newblock \href {http://arxiv.org/abs/hep-th/9411108}
  {\path{arXiv:hep-th/9411108}}, \href
  {http://dx.doi.org/10.1016/0550-3213(95)00111-5}
  {\path{doi:10.1016/0550-3213(95)00111-5}}.

\bibitem{Zanotti:2001yb}
J.~M. Zanotti, et~al., {Hadron masses from novel fat link fermion actions},
  Phys.Rev. D65 (2002) 074507.
\newblock \href {http://arxiv.org/abs/hep-lat/0110216}
  {\path{arXiv:hep-lat/0110216}}, \href
  {http://dx.doi.org/10.1103/PhysRevD.65.074507}
  {\path{doi:10.1103/PhysRevD.65.074507}}.

\bibitem{Kamleh:2004aw}
W.~Kamleh, P.~O. Bowman, D.~B. Leinweber, A.~G. Williams, J.~Zhang, {The fat
  link irrelevant clover overlap quark propagator}, Phys.Rev. D71 (2005)
  094507.
\newblock \href {http://arxiv.org/abs/hep-lat/0412022}
  {\path{arXiv:hep-lat/0412022}}, \href
  {http://dx.doi.org/10.1103/PhysRevD.71.094507}
  {\path{doi:10.1103/PhysRevD.71.094507}}.

\bibitem{Kamleh:2004xk}
W.~Kamleh, D.~B. Leinweber, A.~G. Williams, {Hybrid Monte Carlo with fat link
  fermion actions}, Phys.Rev. D70 (2004) 014502.
\newblock \href {http://arxiv.org/abs/hep-lat/0403019}
  {\path{arXiv:hep-lat/0403019}}, \href
  {http://dx.doi.org/10.1103/PhysRevD.70.014502}
  {\path{doi:10.1103/PhysRevD.70.014502}}.

\bibitem{Kamleh:2001ff}
W.~Kamleh, D.~H. Adams, D.~B. Leinweber, A.~G. Williams, {Accelerated overlap
  fermions}, Phys.Rev. D66 (2002) 014501.
\newblock \href {http://arxiv.org/abs/hep-lat/0112041}
  {\path{arXiv:hep-lat/0112041}}, \href
  {http://dx.doi.org/10.1103/PhysRevD.66.014501}
  {\path{doi:10.1103/PhysRevD.66.014501}}.

\bibitem{Luscher:1984xn}
M.~L{\"u}scher, P.~Weisz, {On-Shell Improved Lattice Gauge Theories},
  Commun.Math.Phys. 97 (1985) 59.
\newblock \href {http://dx.doi.org/10.1007/BF01206178}
  {\path{doi:10.1007/BF01206178}}.

\bibitem{Davies:1987vs}
C.~Davies, G.~Batrouni, G.~Katz, A.~S. Kronfeld, G.~Lepage, et~al., {Fourier
  Acceleration in Lattice Gauge Theories. 1. Landau Gauge Fixing}, Phys.Rev.
  D37 (1988) 1581.
\newblock \href {http://dx.doi.org/10.1103/PhysRevD.37.1581}
  {\path{doi:10.1103/PhysRevD.37.1581}}.

\bibitem{Bonnet:1999mj}
F.~D. Bonnet, P.~O. Bowman, D.~B. Leinweber, A.~G. Williams, D.~G. Richards,
  {Discretization errors in Landau gauge on the lattice}, Austral.J.Phys. 52
  (1999) 939--948.
\newblock \href {http://arxiv.org/abs/hep-lat/9905006}
  {\path{arXiv:hep-lat/9905006}}, \href {http://dx.doi.org/10.1071/PH99047}
  {\path{doi:10.1071/PH99047}}.

\bibitem{Leinweber:1998im}
D.~B. Leinweber, J.~I. Skullerud, A.~G. Williams, C.~Parrinello, {Gluon
  propagator in the infrared region}, Phys.Rev. D58 (1998) 031501.
\newblock \href {http://arxiv.org/abs/hep-lat/9803015}
  {\path{arXiv:hep-lat/9803015}}, \href
  {http://dx.doi.org/10.1103/PhysRevD.58.031501}
  {\path{doi:10.1103/PhysRevD.58.031501}}.

\bibitem{Hollwieser:2008tq}
R.~H{\"o}llwieser, M.~Faber, J.~Greensite, U.~M. Heller, S.~Olejnik, {Center
  Vortices and the Dirac Spectrum}, Phys.Rev. D78 (2008) 054508.
\newblock \href {http://arxiv.org/abs/0805.1846} {\path{arXiv:0805.1846}},
  \href {http://dx.doi.org/10.1103/PhysRevD.78.054508}
  {\path{doi:10.1103/PhysRevD.78.054508}}.

\bibitem{BilsonThompson:2002jk}
S.~O. Bilson-Thompson, D.~B. Leinweber, A.~G. Williams, {Highly improved
  lattice field strength tensor}, Annals Phys. 304 (2003) 1--21.
\newblock \href {http://arxiv.org/abs/hep-lat/0203008}
  {\path{arXiv:hep-lat/0203008}}, \href
  {http://dx.doi.org/10.1016/S0003-4916(03)00009-5}
  {\path{doi:10.1016/S0003-4916(03)00009-5}}.

\bibitem{Moran:2008qd}
P.~J. Moran, D.~B. Leinweber, {Impact of Dynamical Fermions on QCD Vacuum
  Structure}, Phys.Rev. D78 (2008) 054506.
\newblock \href {http://arxiv.org/abs/0801.2016} {\path{arXiv:0801.2016}},
  \href {http://dx.doi.org/10.1103/PhysRevD.78.054506}
  {\path{doi:10.1103/PhysRevD.78.054506}}.

\bibitem{Hollwieser:2014osa}
R.~H{\"o}llwieser, M.~Faber, T.~Schweigler, U.~M. Heller, {Chiral Symmetry
  Breaking from Center Vortices}\href {http://arxiv.org/abs/1410.2333}
  {\path{arXiv:1410.2333}}.

\end{thebibliography}

\end{document}